\documentclass[prb,preprint,superscriptaddress]{revtex4}
\begin{document}
\title{Properties of Gutzwiller wave-functions for multi-band models}
\author{Claudio Attaccalite}
\affiliation{International School for Advanced Studies, SISSA, Via Beirut 2, I-34014 
Trieste, Italy}
\author{Michele Fabrizio}
\affiliation{International School for Advanced Studies, SISSA, Via Beirut 2, I-34014 
Trieste, Italy}
\affiliation{International Center for Theoretical Physics, ICTP, 
Strada Costiera 11, I-34100, Trieste, Italy}
\begin{abstract}
We analyze the Mott transition in multi-band Hubbard models with the inclusion 
of multiplet exchange splittings as it arises in infinite dimensions 
by using the generalized Gutzwiller wave-function 
introduced by B\"unemann, Weber and Gebhard [Phys. Rev. B 
{\bf 57}, 6896 (1998)]. We also present an extension of 
that variational wave-function to account for broken-symmetry solutions, which still 
allows an exact analytical treatment.  Our analysis 
reveals some drawbacks of the variational wave-function, which, 
in our opinion, imply that Gutzwiller-type of wave-functions do not properly characterize  
quasi-particles close to a Mott transition.
\end{abstract}
\maketitle
\section{Introduction}
After so many years since its proposal\cite{Mott}, the Mott transition 
maintains intact its scientific interest, continually 
kept alive by the growing number of materials which, on the track of becoming 
Mott insulators, show rich and fascinating physical properties, among which 
high T$_c$ superconductivity is likely the most spectacular example.

Much of the theoretical effort has been till now devoted towards cuprate-inspired  
mainly single-band models, like the standard Hubbard model or its strongly correlated 
counterpart, the $t-J$ model. Although unsolved questions still remain, a lot has 
been unveiled about the Mott transition in the single-band Hubbard model, especially 
after the development of the so-called Dynamical Mean Field Theory (DMFT)\cite{DMFT}.  
DMFT, which is exact in the limit of infinite coordination lattices, 
represents a quite reliable approach to investigate the on-site dynamical behavior across 
the Mott transition. Yet, in its original version, DMFT is not able to yield 
accurate results concerning inter-site correlations, which are treated in a mean-field like 
fashion, although several extensions has been proposed\cite{ClusterDMFT}. For that  
reason, other approaches has been often adopted to investigate the possible occurrence of 
$d$-wave superconductivity in proximity to the Mott insulating phase of single-band 
Hubbard and $t-J$ models, most of which are more or less explicitly related to 
the Gutzwiller variational technique\cite{Gutzwiller}. That amounts to 
study a variational wave-function consisting of a simple Slater determinant where 
doubly occupied sites are partially or completely projected out. The Gutzwiller wave-function 
can be rigorously handled only numerically in finite dimensions, although  
an approximate analytical scheme to evaluate average values was proposed by Gutzwiller 
himself\cite{Gutzwiller}, 
thereafter called Gutzwiller approximation formula (GAF). Later on, it was realized that  
the GAF is exact in the case of infinite coordination 
lattices\cite{Metzner} and that it provides similar results 
to the slave-boson method within the saddle-point approximation\cite{Metzner,Gebhard1}. 
Moreover, the comparison with the numerical 
treatment of the Gutzwiller wave-function has shown that the GAF is quite accurate 
also in finite dimensions\cite{Gros,Shiba,Gebhard1}.

In spite of the interesting developments achieved by means of the Gutzwiller variational 
wave-function in single-band Hamiltonians, 
especially for what it concerns $d$-wave superconductivity 
in the $t$-$J$ model\cite{GrosSC,Randeria,SorellaPRL}, there have been not many attempts to 
apply the same technique to multi-band models, even though there are many interesting 
strongly correlated materials where orbital degrees of freedom play 
an important role. 
Indeed there exist several analyzes based on multi-band extensions 
of the GAF\cite{Lu,Okabe,Gebhard} 
and of slave-boson technique\cite{Hasegawa}, but a direct comparison 
with more exact results to test the quality of the Gutzwiller wave-function 
is still missing, even though  
DMFT results for multi-band Hubbard models are 
by now available\cite{Rozenberg,Bulla,Florens,Koga,Massimo}.

In this paper, we review the multi-band extension of the Gutzwiller 
wave-function (GWF) proposed by Ref. \onlinecite{Gebhard} 
which has the big advantage of being analytically treatable in the limit of infinite 
dimensions. 
Moreover, we propose a further extension to broken-symmetry phases,  
which maintains the same property of being 
analytically accessible in infinite dimensions without loosing any variational freedom.     
In this limit, we single out some 
peculiar properties of the GWF close to the Mott transition. The comparison 
with exact DMFT results reveals some drawbacks of 
the GWF which may lead to even qualitatively incorrect results.

The plan of the paper is as follows. In Section \ref{The Model} we present 
the generalized GWF introduced in Ref. \onlinecite{Gebhard} for multi-band models 
and calculate the general expression of the variational energy in infinite dimensions. 
In Section III we study the infinite-dimension Mott transition both in the absence and 
in the presence of multiplet exchange splittings, which we discuss 
more in detail in Section IV.  In Section V we present an extension of the 
GWF to symmetry broken phases which we use for a specific two-band model in Section  
VI. Finally, in Section VII we draw some conclusions.

\section{The Model}
\label{The Model}
We consider a multi-band lattice model where, on each lattice site, $N$ 
valence orbitals are available for being occupied by the conduction 
electrons. Since our purpose is to discuss some general features of the 
Gutzwiller wave-function, we will assume the most simple form of 
tight-binding energy with nearest neighbor hopping matrix elements 
diagonal in the orbital index, namely
\begin{equation}
\hat{H}_0 = -\frac{t}{\sqrt{z}}\sum_{\langle ij \rangle} \sum_{a=1}^N \sum_\sigma 
c^\dagger_{i,a\sigma} c^{\phantom{\dagger}}_{j,a\sigma} + H.c.,
\label{H0}
\end{equation}
where $z$ is the lattice coordination, $c^{\phantom{\dagger}}_{i,a\sigma}$ and 
$c^\dagger_{i,a\sigma}$ are, respectively, the annihilation and creation 
operators for an electron at site $i$ with orbital index $a=1,\dots,N$ 
and spin $\sigma=\uparrow,\downarrow$. 
The correlation among the electrons is introduced via two local 
interaction terms: an on-site Hubbard repulsion
\begin{equation}
\hat{H}_U = \frac{U}{2}\sum_i n_i^2,
\label{HU}
\end{equation}
where $n_i = \sum_{a=1}^N \sum_\sigma 
c^\dagger_{i,a\sigma} c^{\phantom{\dagger}}_{i,a\sigma}$ is the 
electron occupation number at site $i$, as well as an exchange splitting 
term 
\begin{equation}
\hat{H}_{J} = \sum_i \sum_{n=0}^{2N} \sum_{\Gamma_n} 
J_{\Gamma_n} \, |i,n,\Gamma_n\rangle\langle i,n,\Gamma_n |.
\label{HJ}
\end{equation}
Here $|i,n,\Gamma_n\rangle$ denotes a multiplet of $n$-electron states at site $i$. 
$\Gamma_n$ has multiplicity  $g_{\Gamma_n}$ such that 
\[
\sum_{\Gamma_n} g_{\Gamma_n} \equiv g_n = 
\left(
\begin{array}{c}
2N\\
n\\
\end{array}
\right),
\]
being the binomial on the right hand side the total number $g_n$ of 
available on-site $n$-electron states. Without loss of generality,  
we assume that $\hat{H}_J$ only splits multiplets at fixed $n$ without affecting their 
center-of-gravity energy, implying $\sum_{\Gamma_n} g_{\Gamma_n} J_{\Gamma_n} = 0$.   

In the absence of interaction, the ground state $|\Phi_0 \rangle$ of (\ref{H0}) 
with an average electron number per site $n_0$ 
consists of $N$ degenerate bands each one at filling $n_0/2N$, namely  
\begin{equation}
|\Phi_0 \rangle = \prod_{a=1}^N \prod_{|\vec{k}|<k_F} 
c^\dagger_{k,a\uparrow}c^\dagger_{-k,a\downarrow}|0\rangle,
\label{Phi_0}
\end{equation}
where 
\[
\frac{2}{V}\sum_{|\vec{k}|<k_F} = \frac{n_0}{N}.
\]
If $0<n_0<2N$ is integer, we expect that, for sufficiently large $U$, 
the correlated ground state of 
$\hat{H}=\hat{H}_0 + \hat{H}_U + \hat{H}_J$ describes a Mott insulator. 
Analogously to 
what has been done for the single-band 
Hubbard model, we would like to have a system which allows to study a 
metal-to-insulator transition into an ideal  
{\sl paramagnetic} Mott insulator, in contrast to a more conventional 
metal-to-band-insulator transition.
For that reason, we further assume that the multiplet exchange 
splitting (\ref{HJ}) is such that, within a perturbation expansion  
upon the uncorrelated ground state $|\Phi_0 \rangle$, 
all self-energy 
diagrams are diagonal in spin and orbital indices and independent of them.  
This does not exclude that spin and/or orbital symmetry may be spontaneously 
broken especially close to 
the Mott transition, as it is known to occur for the single band Hubbard model on a 
bipartite lattice. Yet, in what follows, we will discard such a possibility 
and just discuss an idealized MIT, where neither the metal nor the Mott insulator break 
any of the symmetries of the Hamiltonian.  
We postpone to Section \ref{Symmetry breaking} the analysis 
of spontaneous symmetry breaking nearby the Mott transition.

Having this in mind, we start analyzing the role of the interaction by a 
Gutzwiller variational approach. Namely we search 
for the best variational wave-function of the form 
\begin{equation}
|\Psi_G\rangle = \prod_i \hat{P}_{i,G} |\Phi_0\rangle,
\label{Psi_G}
\end{equation}
where the uncorrelated wave-function is the Slater determinant of Eq. (\ref{Phi_0}) and 
the on-site Gutzwiller projector \cite{Gebhard,Okabe},     
\begin{equation}
\hat{P}_{i,G} = \sum_n \sum_{\Gamma_n} \lambda_{n \Gamma_n}\,  
|i,n,\Gamma_n\rangle\langle i,n,\Gamma_n |, 
\label{PiG}
\end{equation}
tends to go along with 
the local interaction terms $\hat{H}_U$ and $\hat{H}_J$ 
in modifying the relative weights of on-site electronic configurations. 
As shown by Refs. \onlinecite{Gebhard1,Gebhard}, 
there is a certain arbitrariness in the choice 
of the variational 
parameters $\lambda$'s, related to the fact that any transformation 
acting on $|\Phi_0\rangle$ and involving 
operators of which $|\Phi_0\rangle$ is an eigenstate amounts simply to a multiplicative factor. 
In our case, that arbitrariness allows to impose, without losing generality,  
the normalization condition 
\begin{equation}
\langle \Phi_0 | \hat{P}_{i,G}^2 | \Phi_0 \rangle, 
\label{cond-1}
\end{equation}
as well as an additional constraint on the single-particle density matrix  
\begin{equation}
\langle \Phi_0 | \hat{P}_{i,G}\, c^\dagger_{i,a\sigma} 
c^{\phantom{\dagger}}_{i,b\sigma'}\, \hat{P}_{i,G}  | \Phi_0 \rangle = 
\langle \Phi_0 | c^\dagger_{i,a\sigma} 
c^{\phantom{\dagger}}_{i,b\sigma'}  | \Phi_0 \rangle 
= \frac{n_0}{2N}\delta_{ab}\delta_{\sigma\sigma'}, 
\label{cond-2}
\end{equation}
where the last equality derives from our choice of $|\Phi_0\rangle$, Eq. (\ref{Phi_0}). 
   
A formal solution of (\ref{cond-1}) and (\ref{cond-2}) is obtained\cite{Gebhard} by writing 
\begin{equation}
\lambda_{n\Gamma_n}^2 = \frac{P(n,\Gamma_n)}{P^{(0)}(n,\Gamma_n)},
\label{lambda}
\end{equation}
where $P^{(0)}(n,\Gamma_n)$ and $P(n,\Gamma_n)$ represent the occupation probabilities of the 
$n$-electron multiplet $\Gamma_n$  
in the uncorrelated, $|\Phi_0\rangle$, and correlated, $|\Psi_G\rangle$, wave-functions. 
Upon inserting (\ref{lambda}) into (\ref{cond-1}) and (\ref{cond-2}) one obtains 
\begin{equation}
\sum_n \sum_{\Gamma_n} P(n,\Gamma_n) = 1,
\label{normalization}
\end{equation}
the correct normalization for $P(n,\Gamma_n)$, as well as
\begin{equation}
\sum_n \sum_{\Gamma_n} n\, P(n,\Gamma_n) = n_0,
\label{n-conservation}
\end{equation}
namely the condition that the average number of electrons per site coincides with the 
uncorrelated value $n_0$.  
In this representation,   
the correlated probability distribution is the variational quantity which has to be optimized. 

The Eqs. (\ref{cond-1}) and (\ref{cond-2}) imply that, within a perturbation expansion in the 
parameters $\left(1-\lambda_{n\Gamma_n}\right)$'s, 
only more than two fermionic lines can exit from any vertex $\hat{P}_{i,G}^2$ at 
site $i$, see Ref.~\onlinecite{Gebhard}. 
This property simplifies considerably the calculations in the limit of infinite coordination 
lattices, $z\to \infty$ in Eq. (\ref{H0}), 
where one can show that the average variational energy per site \cite{Gebhard}
\begin{eqnarray}
E_G &=& \frac{1}{V}\langle \Psi_G | \hat{H}_0 + 
\hat{H}_U + \hat{H}_J | \Psi_G \rangle \nonumber \\
&=& 
Z\, T_0 + \frac{U}{2}\sum_n \sum_{\Gamma_n} (n-n_0)^2 \, P(n,\Gamma_n) 
+ \sum_n \sum_{\Gamma_n} J_{\Gamma_n} \, P(n,\Gamma_n).
\label{EG}
\end{eqnarray}
$T_0$ is the average value per site of the non-interacting tight-binding Hamiltonian
\[
T_0 = \lim_{z\to\infty} \frac{1}{V} \langle \Phi_0 | \hat{H}_0 | \Phi_0 \rangle,  
\]
which gets reduced by a factor $Z$ through the Gutzwiller projection, and we have introduced 
a chemical potential term which makes the Hubbard interaction minimum at the average $n=n_0$. 
Our choice of $\hat{H}_J$  leads to the following expression 
\begin{eqnarray}
\sqrt{Z} &=& \sum_{n=1}^{2N} \sum_{\Gamma_n,\Gamma_{n-1}} 
\frac{n}{2N} \frac{g_{\Gamma_n} g_{\Gamma_{n-1}}}{g_{n-1}} \left(\frac{n_0}{2N}\right)^{n-1} 
\left(1-\frac{n_0}{2N}\right)^{2N-n} \lambda_{n\Gamma_n}\lambda_{n-1\Gamma_{n-1}}  \nonumber \\
&=& \sum_{n=1}^{2N} \sum_{\Gamma_n,\Gamma_{n-1}}\frac{n}{g_{n-1}} 
\sqrt{\frac{g_{\Gamma_n} g_{\Gamma_{n-1}}}{n_0\left(2N-n_0\right)}}  
\sqrt{P(n,\Gamma_n)P(n-1,\Gamma_{n-1})}.
\label{Z}
\end{eqnarray}

When $\hat{H}_J=0$, namely if nothing in the Hamiltonian 
splits the degeneracy 
among states at fixed $n$, there should exist a variational solution 
in which all configurations at fixed $n$ are 
equally probable, a property owned by the Slater determinant $|\Phi_0\rangle$. In that case, 
a multiplet with degeneracy $g_{\Gamma_{n}}$ occurs with probability  
\begin{equation}
P(n,\Gamma_n) = \frac{g_{\Gamma_n}}{g_n} P(n),
\label{Pn-JH=0}
\end{equation}
being $P(n)$ the probability of a site to be occupied by $n$ electrons irrespective 
of the configuration. Eq. (\ref{Pn-JH=0}) inserted into (\ref{Z}) leads to 
the following simple expression of $Z$ valid for $\hat{H}_J=0$:  
\begin{equation}
\sqrt{Z} = \sum_{n=1}^{2N} \frac{n}{\sqrt{n_0\left(2N-n_0\right)}} 
\sqrt{\frac{g_{n}}{ g_{n-1}}}  
\sqrt{P(n)P(n-1)}.
\label{Z-JH=0}
\end{equation}

\section{Mott transition within the Gutzwiller approach}
\label{Mott transition}
The search for the optimal $P(n,\Gamma)$'s with arbitrary values of the interaction 
parameters is not straightforward. 
However it is still possible to obtain simple analytical 
results close to the Mott transition, which is signaled by a vanishing 
{\sl quasiparticle residue} $Z$. Indeed, when 
$Z\ll 1$, one realizes \cite{Lu,nota} by inspection of Eq. (\ref{EG}) that 
\begin{equation}
P(n,\Gamma_n) \sim Z^{|n-n_0|}.
\label{PvsZ}
\end{equation}
It is therefore justified to assume $P(n,\Gamma_n)=0$ for all $n$'s  
but $n=n_0$ and $n=n_0\pm 1$, which simplifies 
remarkably all calculations. 

In the simple case $\hat{H}_J=0$, which we denote hereafter as $J=0$, 
this approximation leads through Eq. (\ref{Z-JH=0}) to 
\begin{eqnarray}
\sqrt{Z} &\simeq& \frac{n_0}{\sqrt{n_0\left(2N-n_0\right)}} 
\sqrt{\frac{g_{n_0}}{ g_{n_0-1}}}  
\sqrt{P(n_0)P(n_0-1)} \nonumber \\
&& + 
\frac{n_0+1}{\sqrt{n_0\left(2N-n_0\right)}} 
\sqrt{\frac{g_{n_0+1}}{ g_{n_0}}}  
\sqrt{P(n_0+1)P(n_0)}.
\label{Zappx-JH=0}
\end{eqnarray}
The distribution probabilities have to satisfy 
\[
P(n_0-1) = P(n_0+1) \equiv d,
\]
and consequently 
\[
P(n_0) = 1 - 2d.
\]
By inserting those expressions into (\ref{Zappx-JH=0}), one obtains 
\begin{eqnarray}
Z &=& \frac{d(1-2d)}{n_0(2N-n_0)} \left[ 
\sqrt{n_0(2N-n_0+1)} + \sqrt{(n_0+1)(2N-n_0)}\right]^2 \nonumber \\
&& \equiv 
\gamma(N,n_0)\, d(1-2d),
\label{Zappx-JH=0-sol}
\end{eqnarray}
hence
\begin{equation}
E_G = \gamma(N,n_0)\, d(1-2d)\, T_0 + U\, d.
\label{EGappx-JH=0}
\end{equation}
The optimal $d$ which minimizes (\ref{EGappx-JH=0}) is readily found: 
\begin{equation}
d = \frac{T_0\gamma(N,n_0) + U}{4 T_0 \gamma(N,n_0)} 
\equiv \frac{U_c(J=0) - U}{4U_c(J=0)},
\label{dappx-JH=0}
\end{equation}
where 
\begin{equation}
U_c(J=0) = - \gamma(N,n_0) \, T_0  = - 
\frac{1}{2N-n_0} \left[ 
\sqrt{n_0(2N-n_0+1)} + \sqrt{(n_0+1)(2N-n_0)}\right]^2 \frac{T_0}{n_0},
\end{equation}
is the value of the interaction at the Mott transition \cite{Lu}, when the optimal 
$d=0$. The variational energy is therefore 
\begin{equation}
E_G(J=0) = 2\,T_0 \, \gamma(N,n_0)\, d^2 = -\frac{1}{8} 
\frac{\left[U_c(J=0)-U\right]^2}{U_c(J=0)}.
\label{EGapp-JH=0-sol}
\end{equation}

Let us now consider the case $H_{J}\not = 0$. We assume that the exchange 
splitting favors at any given $n$ a particular multiplet of states, which we denote as 
$\Gamma^*_{n}$.
The Mott insulator described by the Gutzwiller wave-function 
is therefore characterized by $P(n_0,\Gamma^*_{n_0}) = 1$ and has 
energy $E_{ins} = J_{\Gamma^*_{n_0}}<0$. 
In order to better understand how the Mott transition occurs when $J\not = 0$, 
it is convenient to consider separately two extreme cases which do not require any 
numerical calculation.

Let us start by considering an hypothetical metallic solution able to smoothly transform 
into the Mott insulator, in which therefore only  
the multiplets $\Gamma^*_n$ favored by $H_J$ are occupied close to the MIT. Since only 
$n_0\pm 1$ and $n_0$ are relevant, one obtains an expression of $Z$ similar to 
(\ref{Zappx-JH=0-sol}) with the only difference that 
\begin{eqnarray*}
&&\gamma(N,n_0) \rightarrow 
\gamma_{eff}(N,n_0) 
\equiv \frac{1}{n_0(2N-n_0)}\left[ 
\sqrt{n_0(2N-n_0+1)}\sqrt{\frac{g_{\Gamma^*_{n_0-1}}}{g_{n_0-1}}
\frac{g_{\Gamma^*_{n_0}}}{g_{n_0}}} \right. \\
&& \left.  + \sqrt{(n_0+1)(2N-n_0)}     
\sqrt{\frac{g_{\Gamma^*_{n_0+1}}}{g_{n_0+1}}
\frac{g_{\Gamma^*_{n_0}}}{g_{n_0}}}\,
\right]^2  < \gamma(N,n_0).
\end{eqnarray*}
The variational energy reads now 
\begin{eqnarray*}
E_G(J\not = 0) &=& \gamma_{eff}(N,n_0)\, d(1-2d)\, T_0  + U\, d \\
&& + \left(J_{\Gamma^*_{n_0-1}}+ J_{\Gamma^*_{n_0+1}}-2J_{\Gamma^*_{n_0}}\right)\, d
+ J_{\Gamma^*_{n_0-1}}.
\end{eqnarray*}
Provided we substitute $\gamma \rightarrow \gamma_{eff}$ and 
\[
U \rightarrow U_{eff} \equiv U 
+ \left(J_{\Gamma^*_{n_0-1}}+ J_{\Gamma^*_{n_0+1}}-2J_{\Gamma^*_{n_0}}\right),
\]
which is the actual Hubbard repulsion measured with respect to the energies of 
the lowest multiplets and not from the centers of gravity,   
the formal solution has the same expression as before. In particular the 
critical interaction at which that metallic phase becomes unstable is now 
\begin{eqnarray}
U_c(J\not=0) &=& - \gamma_{eff}(N,n_0) \, T_0 - 
\left(J_{\Gamma^*_{n_0-1}}+ J_{\Gamma^*_{n_0+1}}-2J_{\Gamma^*_{n_0}}\right) 
\nonumber \\
&=& \frac{\gamma_{eff}(N,n_0)}{\gamma(N,n_0)} U_c (J=0) 
- 
\left(J_{\Gamma^*_{n_0-1}}+ J_{\Gamma^*_{n_0+1}}-2J_{\Gamma^*_{n_0}}\right).
\label{Uc_Jnot0}
\end{eqnarray}
For very small $J$'s, $U_c(J\not=0)$ is shifted down with respect to $U_c(J=0)$ by terms 
of order $|T_0|$, which already suggests that the above solution is not the most energetically 
favorable, although it has the merit to merge smoothly into the insulator.  

Indeed, the best variational solution is actually different from the above one. 
If the $J$'s are very small compared with $T_0$, 
we rather expect in the metallic phase that the probability 
distributions of the multiplets at any $n$ are only slightly modified with respect to  
the $J=0$ case. If this were true, we could still search for a variational solution 
of the same form as for $J=0$, which has an energy given by 
(\ref{dappx-JH=0}) with $d$ as in Eq.(\ref{EGapp-JH=0-sol}).  This solution, however, does 
not converge into the insulating one, 
which has an energy $E_{ins} = J_{\Gamma^*_{n_0}}$. The two energies indeed cross when
\[
-\frac{1}{8} 
\frac{\left[U_c(J=0)-U\right]^2}{U_c(J=0)} = J_{\Gamma^*_{n_0}},
\]
namely when $U=U_*$,  
\begin{equation}
U_* = U_c(J=0) - \sqrt{-8U_c(J=0)J_{\Gamma^*_{n_0}}}.
\label{U_*}
\end{equation}
For small $J$, $U_*$ is larger than $U_c(J\not = 0)$ given 
in Eq. (\ref{Uc_Jnot0}), which suggests not only that the 
$J=0$ metallic solution has lower energy but also that the Mott transition is first order. 
The explicit solution of the variational equations shows that,   
in the optimal metal at small $J$, the probability distributions are indeed 
only slightly modified 
with respect to the $J=0$ case by terms of order $\sqrt{|J/T_0|}$ hence that the 
Mott transition becomes first order as soon as a finite $J$ is introduced.

In conclusion, if the multiplet exchange splitting term $|J|$ is much smaller that the 
uncorrelated bandwidth $W$, then the Gutzwiller variational approach leads to a 
first order phase transition from a metal, slightly modified with respect to the $J=0$ case, 
into a Mott insulator instead dominated by $J$. This transition  is predicted to 
occur when the quasiparticle residue 
\begin{equation}
Z \sim \sqrt{\gamma(N,n_0)\left|\frac{J}{T_0}\right|},
\label{ZsqrtJ}
\end{equation}
and arises because the metallic solution has to pay too much hopping-energy to 
modify the relative weights of the multiplets, a cost which overcomes the 
exchange splitting energy gain. 

A first order phase transition has indeed been found in Ref.~\onlinecite{Gebhard}
by explicit numerical minimization of the variational energy in a two-band model, and 
agrees with linearized-DMFT results obtained in Ref.~\onlinecite{Bulla} on a similar model. 
Moreover, in the same Ref.~\onlinecite{Bulla} it is shown that the first 
order character reinforces with increasing  
exchange splitting strength $J$ from $J=0$, while it weakens with further increasing $J$
above some intermediate value. This also agrees with our above results. Indeed, when 
$J$ gets so large to make $U_c(J\not =0)$, Eq. (\ref{Uc_Jnot0}), 
greater than $U_*$, Eq. (\ref{U_*}), we expect the Mott transition to turn again into a 
second order one.  That anyway requires a substantial $|J| \sim |T_0|$.

\section{Drawbacks of the Gutzwiller wave-function}
\label{Weakness}

In the previous Section we have identified the qualitative behavior across the Mott 
transition for multi-band Hubbard models within the Gutzwiller variational approach. 
In particular our analysis has been performed in the limit of infinite coordination 
lattices, where the Gutzwiller wave-function is able to describe a Mott transition. 
In the same limit, exact 
results can be obtained by DMFT\cite{DMFT}, which allows a direct comparison 
hence a test on the quality of the Gutzwiller wave-function. 

As we previously said, upon approaching the Mott transition with a 
Gutzwiller wave-function for multi-band models, the on-site charge configurations 
with a number of electrons $n$ different from the average one $n_0$ get suppressed 
approximately like $Z^{|n-n_0|}$, where $Z$ is the quasiparticle residue. 
It is usually assumed that the GWF in the metallic phase of a single 
band model provides a faithful description of what Landau quasiparticles do, while it gives a 
very poor characterization of the underlying insulating component. If this property holds 
in multi-band models too, then the quasiparticle gas should be characterized 
by a $P(n) \sim Z^{|n-n_0|}$. This behavior is not displayed by exact DMFT results, 
which rather indicate a quasiparticle probability distribution 
$P(n) \sim Z$ (see Ref.~\onlinecite{Massimo}). The origin of this disagreement 
can be easily traced back.  

The main reason why the Gutzwiller wave-function gives a poor 
description of the Mott insulator in the single band Hubbard model at half-filling  
is that it can not account for virtual excitations into unfavorable charge configurations. 
Those virtual processes are responsible of the finite  
average value of the double occupancy $\langle n_\uparrow n_\downarrow\rangle $ 
in the insulating phase even in infinite dimensions. 
Indeed $\langle n_\uparrow n_\downarrow\rangle $ 
as function of $U$ displays a discontinuity 
in the slope across the Mott transition in infinite dimensions\cite{DMFT}. 
The singular part, which vanishes at the MIT, is attributed 
to the quasiparticles and it is qualitatively reproduced by the behavior 
of the double occupancy as obtained through the Gutzwiller wave-function, supporting the 
common belief that this wave-function does correctly capture quasiparticle properties.   

In a multi-band Hubbard model the situation is different. If we just 
consider the occupation probability $P(n_0\pm 1)$ of 
$(n_0\pm 1)$-charge configurations, 
we expect across the MIT a behavior similar to the double occupancy 
in the single-band Hubbard model, still compatible with the GWF. 
Let us instead consider the occupation probability $P(n_0\pm 2)$. In the GWF 
close to the Mott transition, those configurations get suppressed 
like $Z^2$. In reality, virtual processes from the more advantageous 
$n_0$ and $(n_0\pm 1)$ charge configurations imply first of all that 
$P(n_0\pm 2)$ is finite in the insulating phase too, and secondly that the singular 
$(n_0\pm 2)$ quasiparticle contribution still linearly vanishes across the MIT, 
unlike what it is found by the GWF. 
That disagreement is more profound than what it would seem to be. A quasiparticle 
probability distribution of the Gutzwiller type, namely $P(n)\sim Z^{|n-n_0|}$, 
suggests that quasiparticles remain more strongly interacting 
than implied by the true behavior $P(n)\sim Z$, even after the Hubbard 
side-bands are well formed.  This indicates that, unlike what happens in 
the single-band Hubbard model, the multi-band GWF is not fully adequate to 
capture quasiparticle properties.   

The second failure of the Gutzwiller variational approach regards the onset of the 
first order phase transition, which is predicted to occur when $Z\sim \sqrt{|J|}$, for 
$|J|\ll |T_0|$. 
It also originates by the lack of virtual processes. If $J=0$, the Gutzwiller wave-function 
leads to a Mott insulator with a finite entropy, related to the finite number of 
degenerate on-site electronic configurations with $n_0$ electrons. 
This state has an infinite susceptibility to 
a term $\hat{H}_J$ which splits that degeneracy,   
with an energy gain linear in $J$. This result is obviously wrong. The super-exchange 
terms generated by virtual processes into unfavorable charge configurations lead to 
finite susceptibilities even in the Mott insulator. It implies that the actual 
energy gain is quadratic in $J$ so that the Mott transition is either second order or 
weakly first order, in this case occurring when $Z\sim {|J|}$. 

Indeed, this aspect is not peculiar to a multi-band model but also occurs 
in a single band model in the presence of a magnetic field $B$ which splits 
spin-up singly occupied sites from spin-down ones. Also in that case the Gutzwiller approach 
would predict a first order transition when $Z\sim \sqrt{B}$, while in reality, being the 
magnetic susceptibility finite\cite{DMFT},  
the transition occurs at smaller $Z$, see Ref.~\onlinecite{Georges}, 
likely when $Z\sim B$.     

Yet those defects of the Gutzwiller wave-function might not    
affect qualitatively the physical behavior in the most common situations where 
the multiplet exchange splitting term leads to the conventional Hund's rules, namely 
favors high spin and angular momentum configurations. These states are 
usually representable by 
a Slater determinant hence are accessible by a mean-field approach which can be 
improved by the Gutzwiller projector. However there may be less conventional but 
still interesting cases where the multiplet exchange splitting favors low degeneracy 
states, which are 
not Slater determinants hence unaccessible to mean field theories. Here the Gutzwiller 
wave-function might be inadequate to describe the Mott transition mainly because 
it is unable to 
access the interesting region where the metallic kinetic energy gain $\sim Z|T_0|$ 
competes with the exchange splitting $\sim J$.  

The inability of the GWF to describe phases which are not accessible by Hartree-Fock, like 
an {\sl ideal} Mott insulator,  
is a well known fact in finite dimensions\cite{Shiba}. However, it is common opinion  
that the GWF is instead able to yield a correct description 
also for Hartree-Fock-unlike physical situations in infinite dimensions. Even more, 
it is believed that the infinite-dimensional picture provided by the GWF 
is qualitatively right even in finite dimensions. The above discussion suggests that 
this common belief is partly wrong and that an improvement of the GWF towards 
including virtual inter-site processes is necessary\cite{nota-b}.

\section{Gutzwiller wave-functions for symmetry broken phases}
\label{Symmetry breaking}

In spite of its 
appealing features, an {\sl ideal} Mott insulator at zero temperature is unlikely to exist, 
especially if it has huge degeneracy.
Commonly one expects a symmetry broken phase to occur at low temperature. 
For instance, in a single-band model at half-filling, the ideal Mott insulator has an infinite 
spin degeneracy which is likely reduced at low temperature by some magnetic ordering. 
Therefore, even though any mean-field type of approach (including more sophisticated ones based 
on Density Functional Theory) can only stabilize correlated insulators in broken-symmetry 
phases, namely can only describe band-insulators thus hiding the basic phenomena 
leading to a Mott insulator, yet they often provide a 
faithful description of the low temperature physics. 

In this situation, the Gutzwiller variational approach 
should still be useful to improve Hartree-Fock approximation.
That would amount to search for the best wave-function of the form  
\begin{equation}
| \Psi_G(\Delta) \rangle = \hat{P}_G \, | \Phi(\Delta_0) \rangle 
= \prod_i \hat{P}_{i,G} \, | \Phi(\Delta_0) \rangle,
\label{PsiG-broken}
\end{equation}
with $\hat{P}_{i,G}$ still given by Eq. (\ref{PiG}), and 
where $| \Phi(\Delta_0) \rangle$ is a symmetry-broken uncorrelated trial 
wave-function 
with single-particle order parameter $\Delta_0$. In general, after Gutzwiller projection, 
the correlated wave-function will have a different order parameter $\Delta$. 
This implies that the average values of the single-particle density 
matrix over $| \Psi_G(\Delta) \rangle$ and $| \Phi(\Delta_0) \rangle$ do not coincide. 
This does not obviously represent a problem for a numerical treatment, whereas  
would seem to prevent the use of the method developed in Ref. \onlinecite{Gebhard} for 
analytically evaluating the average values in infinite dimensions. In fact that method 
relies on the possibility of constructing a GWF with the same average value of the 
single particle density matrix as the uncorrelated wave-function. We could still impose 
that condition for the wave-function defined by (\ref{PsiG-broken}), but that would reduce 
the variational freedom.   

In this Section we 
present a simple extension of the GWF to account for symmetry broken phases while leaving the 
property of being analytically treatable in infinite dimensions without any variational loss.  
We start by noticing that there always exists a non unitary operator 
$\hat{U} = \prod_i \hat{U}_i$ such that its action  
\begin{equation}
\hat{U} \, | \Phi(\Delta_0) \rangle = | \Phi(\Delta) \rangle,
\label{def:U}
\end{equation}
leads to a trial wave-function $| \Phi(\Delta) \rangle$ of the same form as 
$| \Phi(\Delta_0) \rangle$ but with the same average value of the 
single-particle density matrix as the Gutzwiller projected $| \Psi_G(\Delta) \rangle$, 
hence the same order parameter $\Delta$. Therefore, 
\begin{equation}
| \Psi_G(\Delta) \rangle = \hat{P}_G \, | \Phi(\Delta_0) \rangle
= \hat{P}_G \, \hat{U} \, | \Phi(\Delta) \rangle 
\equiv \hat{{\rm P}}_{G} | \Phi(\Delta) \rangle,
\label{PsiG-broken-bis}
\end{equation}
which implies that, provided we substitute  
\begin{equation}
\hat{P}_G \rightarrow \hat{{\rm P}}_G = \hat{P}_G \, \hat{U},
\label{PG-notH}
\end{equation}
we can still search without loss of generality for a variational wave-function 
\[
| \Psi_G(\Delta) \rangle = \hat{{\rm P}}_G \, | \Phi(\Delta) \rangle,
\]
where the average single-particle density matrix stays unchanged after Gutzwiller projection.   
The cost is that we must work with a Gutzwiller projector as given by (\ref{PG-notH}),  
which is in general neither diagonal in the multiplets which appears in $\hat{H}_J$, 
Eq. (\ref{HJ}), nor hermitean. 

Actually we can identify two distinct situations which may occur, one of them  
being already included in the formalism developed by Ref. \onlinecite{Gebhard}. If the 
exchange splitting $J$ favors a degenerate atomic configuration, then we reasonably expect that 
in the true ground state of the lattice only one of the degenerate states will be occupied on a 
given site, eventually changing from site to site. There the order parameter corresponds 
locally to a conserved quantity of the atomic Hamiltonian, for instance the $z$-component of 
the on-site spin, which leads to a 
generalized Gutzwiller projector $\hat{{\rm P}}_G$ still hermitean and diagonal. 
Let us consider as a simple example 
a one band model. The local Gutzwiller projector is in general 
\[
\hat{P}_{i,G} = \lambda_0\, |i,0\rangle\langle i, 0| 
+ \lambda_2\, |i,2\rangle\langle i,2|  
+ \lambda_1\,\left[ \, |i,\uparrow\rangle\langle i,\uparrow| 
+ |i,\downarrow \rangle\langle i,\downarrow|\, \right],
\]
where $|i,0(2)\rangle$ denote the empty or doubly-occupied site $i$, while 
$|i,\sigma\rangle$ the singly occupied site with spin $\sigma$. We assume that the uncorrelated 
wave-function is magnetically ordered. Then (\ref{def:U}) may be constructed  
by local operators 
\[
\hat{U}_i = {\rm e}^{2 \alpha_i \, \hat{S}^z_{i}},
\]
where $\hat{S}^z_i$ is the $z$-component of the spin operator at site $i$.  We find  
that 
\[
\hat{{\rm P}}_{i,G} = \hat{P}_{i,G}\, \hat{U}_i = 
\lambda_0\, |i,0\rangle\langle i, 0| 
+ \lambda_2\, |i,2\rangle\langle i,2| + 
\lambda_1\, {\rm e}^{\alpha_i}\,
|i,\uparrow\rangle\langle i,\uparrow| 
+ \lambda_1\, {\rm e}^{-\alpha_i}\,
|i,\downarrow \rangle\langle i,\downarrow|.
\]
Indeed the modified Gutzwiller projector is still diagonal and hermitean, although 
there appear different variational parameters for spin up and down components. This additional 
degree of freedom gets fixed once we require that the order parameters of the 
correlated and uncorrelated wave-functions coincide. The above type of GWF is actually 
a particular case of the generalized GWF's introduced in Ref.~\onlinecite{Gebhard}.  
 
We can however envisage a different situation where the uncorrelated wave-function 
$|\Phi_0\rangle$ has an order parameter which does not correspond locally to a conserved 
quantity. There a modified Gutzwiller projector $\hat{{\rm P}}_G$ is unavoidably 
off-diagonal and non-hermitean. Let us discuss an over-simplified example. We assume that 
the exchange term leads to a local Gutzwiller projector of the form (we drop the site label)
\[
\hat{P}_G = \lambda_a\, |a\rangle\langle a| + \lambda_b\, |b\rangle\langle b|,
\]
and moreover that the uncorrelated wave-function has an order parameter 
identified by non-zero matrix elements
\[
\langle \Phi_0 |a\rangle\langle b |\Phi_0\rangle  
= \langle \Phi_0 |b\rangle\langle a |\Phi_0\rangle.
\]
The transformation $\hat{U}$ can be now taken of the form 
\[
\hat{U} = \alpha\,\left(|a\rangle\langle a| + |b\rangle\langle b|\right)  
+ \beta\,\left(|a\rangle\langle b| + |b\rangle\langle a|\right).
\]
We readily find that 
\[
\hat{{\rm P}}_G = \hat{P}_G\, \hat{U} = 
\alpha\, \lambda_a\, |a\rangle\langle a|
+ \alpha\, \lambda_b\, |b\rangle\langle b| 
+ \beta\, \lambda_a\, |a\rangle\langle b|
+ \beta\, \lambda_b\, |b\rangle\langle a|,
\]
indeed containing off-diagonal terms and evidently non-hermitean. This is a novel situation 
which we are going to discuss more in detail in a particular example.  

We conclude this Section by pointing out that the non-hermitean 
character plays a crucial role only if the ordering involves just the quasiparticles, while 
it is essentially irrelevant when both the quasiparticles and the Mott-Hubbard side-bands 
contribute to the order parameter. In the following section we analyse a two-band model 
where both cases may appear. 

\section{A two-band model study} 
\label{A two-band model study}
We consider a two-band Hubbard model 
described by the Hamiltonian (\ref{H0}) and (\ref{HU}) where the orbital index $a=1,2$. 
Besides the local spin-density operators
\[
\vec{S}_i = \frac{1}{2} \sum_{a=1}^2 \sum_{\alpha\beta} 
c^\dagger_{i, a \alpha} \vec{\sigma}_{\alpha\beta} 
c^{\phantom{\dagger}}_{i,a \beta},
\]
where $\hat{\sigma}_x$, $\hat{\sigma}_y$ and $\hat{\sigma}_z$ are the Pauli matrices, 
we introduce orbital pseudo-spin operators
\begin{equation}
\vec{T}_i = \frac{1}{2} \sum_{a,b=1}^2 \sum_{\sigma} 
c^\dagger_{i, a \sigma} \vec{\tau}_{ab} 
c^{\phantom{\dagger}}_{i,b \sigma},
\label{T}
\end{equation}
where the Pauli matrices $\hat{\tau}$'s act on the orbital indices. 
The hopping and Hubbard terms, (\ref{H0}) and (\ref{HU}), have a very large SU(4) symmetry. 
Having in mind common physical realizations, like e.g. $d$-orbitals of $e_g$ symmetry,  
we assume that SU(4) is lowered down to 
a spin SU(2) times an orbital O(2) by the exchange term, which can therefore be written as 
\begin{equation}
\hat{H}_J = \sum_i J_S \, \vec{S}_i\cdot \vec{S}_i 
+ J_T \, \vec{T}_i\cdot\vec{T}_i -3\left(J_S+J_T\right)\left(T_i^z\right)^2.
\label{Exe:H_J}
\end{equation}
The above $\hat{H}_J$ just splits the on-site configurations with two-electrons. 
There are six of those states on a given site $i$, a spin-triplet orbital-singlet 
$|i,n=2;S=1,S_z;T=0 \rangle$, which we denote hereafter 
as $|i,2,t\rangle$, and a spin-singlet orbital-triplet, 
for which we use the short notations 
$|i,n=2;S=0;T=1,T_z=0\rangle\equiv |i,2,0\rangle$ and 
$|i,n=2;S=0;T=1,T_z=\pm 1\rangle \equiv |i,2,\pm\rangle$. In this subspace, $\hat{H}_J$ 
has the form 
\begin{equation}
\hat{H}_J = \sum_i\,  2\,J_S\, |i,2,t\rangle\langle i,2,t| 
+ 2\,J_T \, |i,2,0\rangle\langle i,2,0| - \left(3\,J_S+J_T\right)\, 
|i,2,\pm \rangle\langle i,2,\pm|.
\label{Exe:HJ}
\end{equation}
The standard Hund's rules correspond to $-J_T<J_S<-5\,J_T/6<0$, when the spin-triplet 
orbital-singlet configuration has the lowest energy, followed by the spin-singlet 
orbital doublet with $T_z=\pm 1$.  
In this case the ideal Mott insulator at half-filling, $n_0=2$,  
represents localized spin-1 
moments which should order at low enough temperature to freeze out 
the spin-entropy. On general grounds one expects that the magnetic ordering in the insulator  
should contaminate the nearby metallic phase so that, as $U$ increases from weak coupling, 
first a transition from a paramagnetic into a magnetic metal should occur, followed 
by a Mott transition into a magnetically ordered insulator. 
Even a mean field approach is in principle able to reproduce the above scenario.  
In this situation, as we discussed before, 
the Gutzwiller-projected  Hartree-Fock wave-function does improve the mean-field solution, 
providing a better physical description. 
In Appendix I we analyse in detail the case in which a bipartite lattice stabilizes 
an antiferromagnetic ordering. 
       
Less conventional is the situation $J_T<-|J_S|$, where the non-degenerate 
spin-singlet with $T_z=0$, namely 
\begin{equation}
|i,2;S=0;T=1,T_z=0 \rangle = 
\frac{1}{\sqrt{2}}\left(c^\dagger_{i,1\uparrow}c^\dagger_{i,2\downarrow}
+ c^\dagger_{i,2\uparrow}c^\dagger_{i,1\downarrow}\right) |0\rangle,
\label{S=0-state}
\end{equation}
is the lowest energy configuration. 
That would be for instance the case of two Hubbard models (two-chains, 
two-planes, etc...), coupled by an antiferromagnetic exchange. 
Here the large $U$ Mott insulator with $n_0=2$ describes a collection of on-site singlets, a local version 
of a valence-bond (VB) insulator. Since it is not degenerate and fully gaped, 
we expect the VB insulator to be stable at large $U$ against any spin and/or 
orbital order. Just to avoid unessential complications, we assume 
that the lattice is sufficiently frustrated to prevent any spin/orbital ordering at 
any $U$.
This situation is far less trivial than the previous one. 
In fact, being the Mott insulator not describable by a single Slater determinant, it is 
inherently unreachable by any mean-field approach, which necessarily leads to some 
kind of ordered state. According to our previous discussion, we expect in this case that 
also the GWF is unable to provide a faithful description of the Mott transition.   

As we showed in Section \ref{Mott transition}, the GWF without any symmetry breaking would 
undergo a first order metal-insulator transition when the quasiparticle residue 
\[
Z \simeq \sqrt{\frac{J_T}{6T_0}},
\]
namely when $U=U_*$ being 
\[
U_* = -6 T_0 - 4\sqrt{6\, T_0 \, J_T}.
\]
We also argued that this result is wrong since a metallic phase is able to enter the regime 
in which $Z\sim |J_T/T_0|$. Let us now check whether there exists 
a better broken-symmetry metallic solution. 
Indeed, even if lattice frustration prevents spin/orbital order, there is still 
a broken-symmetry GWF which might in principle compete with the above metallic solution.

\subsection{Superconducting Gutzwiller wave-function}
When $U=0$, the multiplet exchange term  
(\ref{Exe:HJ}) favors a BCS Hartree-Fock wave-function with 
the $s$-wave order parameter 
\begin{equation}
\langle \Phi_{BCS}(\Delta_0)|c^\dagger_{i,1\uparrow}c^\dagger_{i,2\downarrow}
|\Phi_{BCS} (\Delta_0)\rangle 
= \langle \Phi_{BCS}(\Delta_0) | c^\dagger_{i,2\uparrow}c^\dagger_{i,1\downarrow} 
| \Phi_{BCS}(\Delta_0) \rangle 
\equiv \Delta_0\leq \frac{1}{2}.
\end{equation}
When $\Delta_0\to 1/2$, the doubly occupied sites in the spin-triplet, $|2,t\rangle$,    
or in the doublet of spin-singlets, $|2,\pm\rangle$, configurations 
get suppressed by a factor $(1-2\Delta_0)^2$ with respect to  
the $T_z=0$ spin-singlet, $|2,0\rangle$.
Similarly, the probability of singly, $|1\rangle$, or triply-occupied, $|3\rangle$, 
sites vanish like 
$(1-2\Delta_0)$. This suggests that by Gutzwiller projecting out sites with zero and four 
electrons, $|0\rangle$ and $|4\rangle$, respectively, through the variational wave-function 
\begin{equation}
|\Psi_G(\Delta) \rangle = \hat{P}_G \, | \Phi_{BCS}(\Delta_0)\rangle,
\label{Exe:GBCSwf}
\end{equation}
one might indeed smoothly connect to the VB insulating state, with 
$(1-2\Delta_0)$ playing the role of $Z$. However, even though the uncorrelated 
BCS wave-function has a large order parameter $\Delta_0\sim 1/2$, the correlated 
$|\Psi_G\rangle$ should have a much smaller one, $\Delta \sim Z \Delta_0$, since 
only quasiparticles are involved in superconductivity. 
In such a case we are therefore obliged to implement the non-hermitean Gutzwiller projector 
in order to get analytical results for large coordination 
lattices. 
In other words, we shall work with a Gutzwiller wave-function of the same form as 
(\ref{Exe:GBCSwf}) and impose that $|\Psi_G \rangle$ and  $| \Phi_{BCS}\rangle$ 
have the same order parameter $\Delta$ through a non-hermitean $\hat{{\rm P}}_G$, 
see Eq. (\ref{PG-notH}). Since this is a novel situation in the Gutzwiller variational 
approach, we prefer to describe it in detail.
  
In order to simplify the analysis at half-filling, we assume that 
the Hamiltonian has particle-hole symmetry and 
search for solutions which do not break it. This guarantees that there are still 
two conditions we can impose without 
losing variational freedom: an overall normalization, (\ref{cond-1}),  
and a particle-hole symmetry constraint. 

To accomplish this job, it is convenient to work not in the original electron 
basis but in the {\sl natural} basis where the on-site single-particle density 
matrix is diagonal. This is done by the following unitary transformation, which is valid at 
half-filling $n_0=2$,  
\[
\left(
\begin{array}{c}
c^{\phantom{\dagger}}_{i,1(2)\uparrow} \\
c^\dagger_{i,2(1)\downarrow} \\
\end{array}
\right)    = 
\frac{1}{\sqrt{2}}\left(
\begin{array}{lr}
1 & -1 \\
1 & 1 \\
\end{array}
\right) \, 
\left(
\begin{array}{c}
a^{\phantom{\dagger}}_{i,1(2)\uparrow} \\
a^\dagger_{i,2(1)\downarrow} \\
\end{array}
\right).
\]
In the natural basis the only non-zero on-site average is  
\[
\langle \Phi_{BCS}(\Delta) | a^\dagger_{i,a\sigma} a^{\phantom{\dagger}}_{i,a\sigma}
| \Phi_{BCS}(\Delta)\rangle = \frac{1}{2} - \Delta.
\]
In order to distinguish the local configurations in the natural from the original basis 
we will denote the former ones as $|\bar{n},\Gamma_{\bar{n}}\rangle$. 
The most general local Gutzwiller projector is in this case 
\begin{equation}
\hat{{\rm P}}_{i,G} = \sum_{\bar{n}}\sum_{\Gamma_{\bar{n}}} 
\lambda_{\bar{n}\Gamma_{\bar{n}}} \, 
|i,\bar{n},\Gamma_{\bar{n}}\rangle\langle i,\bar{n},\Gamma_{\bar{n}} | 
+ \lambda_{\bar{0}\bar{4}}\, |i,\bar{0}\rangle\langle i,\bar{4} |
+ \lambda_{\bar{4}\bar{0}}\, |i,\bar{4}\rangle\langle i,\bar{0} |.
\label{Exe:PGBCS}
\end{equation}
The last two terms are the only possible off-diagonal elements 
at half-filling when particle-hole symmetry holds. 
The normalization condition and the conservation of the single-particle density-matrix  
lead to the following parametrization of the $\lambda$'s for 
$\bar{n}\not = 0,4$
\[
\lambda_{\bar{n}\Gamma_{\bar{n}}}^2 = \frac{P(\bar{n},\Gamma_{\bar{n}})}
{P^{(0)}(\bar{n},\Gamma_{\bar{n}})},
\]
where $P(\bar{n},\Gamma_{\bar{n}})$ and $P^{(0)}(\bar{n},\Gamma_{\bar{n}})$ are 
the correlated and uncorrelated occupation probabilities in the natural basis. 
For $\bar{n}=0,4$ we have instead 
\begin{eqnarray}
\lambda_{\bar{0}}^2 P^{(0)}(\bar{0})  
+ \lambda_{\bar{0}\bar{4}}^2 P^{(0)}(\bar{4}) &=& P(\bar{0}), \label{Pbar0}\\  
\lambda_{\bar{4}\bar{0}}^2 P^{(0)}(\bar{0})  
+ \lambda_{\bar{4}}^2 P^{(0)}(\bar{4}) &=& P(\bar{4}) \label{Pbar4},\\
\lambda_{\bar{0}}\lambda_{\bar{4}\bar{0}} P^{(0)}(\bar{0})  
+ \lambda_{\bar{0}\bar{4}}\lambda_{\bar{4}} 
P^{(0)}(\bar{4}) &=& A(\bar{0};\bar{4}) = A(\bar{4};\bar{0}) \label{Abar04},
\end{eqnarray}
where we introduce the transition amplitudes 
\[
A(\bar{n},\Gamma_{\bar{n}};\bar{m},\Gamma_{\bar{m}}) 
= \langle \Psi_G  |i,\bar{n},\Gamma_{\bar{n}}\rangle\langle 
i,\bar{m},\Gamma_{\bar{m}}| \Psi_G\rangle.
\]

The occupation probabilities in the natural basis are related to those in the 
original one through 
 \begin{equation}
\begin{array}{lcl}
P(\bar{0}) &=& \frac{1}{2}\left(P(0)+P(2,0)+A(0;4)\right) 
+ \sqrt{2}A(0;2,0), \\
P(\bar{1}) &=& P(1) + A(1;3), \\
P(\bar{2},0) &=& P(0) - A(0;4), \\
P(\bar{2},\pm) &=& P(2,\pm),\\
P(\bar{2},t) &=& P(2,t), \\
P(\bar{3}) &=& P(1) - A(1;3), \\
P(\bar{4}) &=& \frac{1}{2}\left(P(0)+P(2,0)+A(0;4)\right) 
- \sqrt{2}A(0;2,0), \\
A(\bar{0};\bar{4}) &=& \frac{1}{2}\left(P(0)-P(2,0)+A(0;4)\right),\\
\end{array}
\label{Pbar-P}
\end{equation}
where we have used the fact that, by particle-hole symmetry, 
$P(1)=P(3)$ and $P(0)=P(4)$. The order parameter $\Delta$ is given by
\[
2\Delta = 2\sqrt{2} A(02+) + A(13),
\]
and the following inequalities should be verified
\begin{eqnarray*}
2A(0;2,0)^2 &\leq& P(2,0)\left[P(0)+A(0;4)\right],\\
A(0;4) &\leq& P(0),\\
A(1;3) &\leq& P(1).
\end{eqnarray*}

The big advantage in working with natural orbitals is that the 
$Z$ reduction factor has the same expression as in (\ref{Z}), namely 
\begin{equation}
\sqrt{Z(\Delta)} = \sum_{\bar{n}=1}^{4} \sum_{\Gamma_{\bar{n}},\Gamma_{\bar{n}-1}} 
\frac{\bar{n}}{4} 
\frac{g_{\Gamma_{\bar{n}}} g_{\Gamma_{\bar{n}-1}}}{g_{\bar{n}-1}} 
\left(\frac{1}{2} - \Delta \right)^{\bar{n}-1} 
\left(\frac{1}{2} + \Delta\right)^{4-\bar{n}} 
\lambda_{\bar{n}\Gamma_{\bar{n}}}\lambda_{\bar{n}-1\Gamma_{\bar{n}-1}}.
\label{Exe:BCSZ}
\end{equation}
Once we know how to relate the parameters $\lambda$'s and $Z$ to the variational 
occupation probabilities 
defining the correlated and uncorrelated wave-functions, 
we can solve the most general variational problem by minimizing the energy functional 
\begin{eqnarray}
E_G(\Delta) = Z(\Delta)\, T_0(\Delta) 
+ U\, \left[4P(0) + P(1)\right] \nonumber \\
+ 2J_S\, P(2,t) + 2J_T\, P(2,0) -(3J_S+J_T)\, P(2,\pm),
\label{BCS-EG}
\end{eqnarray} 
where 
\[
\langle \Phi_{BCS}(\Delta) | \hat{H}_0 | \Phi_{BCS}(\Delta) \rangle = T_0(\Delta).
\]

For the sake of clarity, we present here an analysis based on 
the following parametrization of the $\lambda$'s in Eq. (\ref{Exe:PGBCS}), 
although it has less variational freedom: 
\begin{equation}
\begin{array}{lcl}
\lambda_{\bar{0}} &=& \frac{1}{2}\left(\lambda_{0} +\lambda_{20} \right) 
\sqrt{\frac{P^{(0)}_{\Delta_0}(\bar{0})}{P^{(0)}_{\Delta}(\bar{0})}},   \\
\lambda_{\bar{4}} &=& \frac{1}{2}\left(\lambda_{0} + \lambda_{20} \right) 
\sqrt{\frac{P^{(0)}_{\Delta_0}(\bar{4})}{P^{(0)}_{\Delta}(\bar{4})}},\\   
\lambda_{\bar{0}\bar{4}} &=& \frac{1}{2}\left(\lambda_{0}- \lambda_{20} \right) 
\sqrt{\frac{P^{(0)}_{\Delta_0}(\bar{4})}{P^{(0)}_{\Delta}(\bar{4})}}, \\   
\lambda_{\bar{4}\bar{0}} &=& \frac{1}{2}\left(\lambda_{0}-\lambda_{20} \right) 
\sqrt{\frac{P^{(0)}_{\Delta_0}(\bar{0})}{P^{(0)}_{\Delta}(\bar{0})}},\\   
\lambda_{\bar{1}} &=&  \lambda_{1} 
\sqrt{\frac{P^{(0)}_{\Delta_0}(\bar{1})}{P^{(0)}_{\Delta}(\bar{1})}},\\   
\lambda_{\bar{3}} &=&  \lambda_{1} 
\sqrt{\frac{P^{(0)}_{\Delta_0}(\bar{3})}{P^{(0)}_{\Delta}(\bar{3})}},\\   
\lambda_{\bar{2}0} &=&  \lambda_{0} 
\sqrt{\frac{P^{(0)}_{\Delta_0}(\bar{2},0)}{P^{(0)}_{\Delta}(\bar{2},0)}}, \\  
\lambda_{\bar{2}t} &=&  \lambda_{2t} 
\sqrt{\frac{P^{(0)}_{\Delta_0}(\bar{2},t)}{P^{(0)}_{\Delta}(\bar{2},t)}},\\   
\lambda_{\bar{2}\pm} &=&  \lambda_{2\pm} 
\sqrt{\frac{P^{(0)}_{\Delta_0}(\bar{2},\pm)}{P^{(0)}_{\Delta}(\bar{2},\pm)}}.\\
\end{array}
\label{lambda-parametrization}
\end{equation}   
As before $\Delta_0$ is the order parameter of the uncorrelated BCS wave-function while 
$\Delta$ is the true order parameter after Gutzwiller projection, see Eq. (\ref{Exe:GBCSwf}). 
$P^{(0)}_{\Delta_0}(\bar{n},\Gamma_{\bar{n}})$ 
and $P^{(0)}_{\Delta}(\bar{n},\Gamma_{\bar{n}})$ are the distribution probabilities 
in the natural basis for the BCS wave-functions with order parameter $\Delta_0$ 
and $\Delta$, respectively. They are explicitly written in the Appendix, 
Eq. (\ref{Exe:BCSP_0}). 

The normalization condition as well as the conservation of the single particle 
density matrix imply that 
\begin{equation}
\lambda_{n\Gamma_n}^2= \frac{P(n,\Gamma_n)}{P^{(0)}_{\Delta_0}(n,\Gamma_n)},
\label{lambda-BCS}
\end{equation}
where $P^{(0)}_{\Delta_0}(n,\Gamma_n) = 
\langle \Phi_{\Delta_0} |i;n,\Gamma_n\rangle 
\langle i;n,\Gamma_n|\Phi_{\Delta_0}\rangle$ is the occupation probability 
for configurations in the original electronic basis within the uncorrelated 
BCS wave-function with large order parameter $\Delta_0$, see Eq. (\ref{App:P_0-original}), 
while $P(n,\Gamma_n)$ is the same quantity for the correlated wave-function. 
Eq. (\ref{lambda-BCS}) is the most natural generalization 
of (\ref{lambda}) to a broken-symmetry  phase, which is the reason why 
we have chosen the above parametrization. The true order parameter $\Delta$ is 
defined through 
\begin{eqnarray}
2\Delta &=& P^{(0)}_{\Delta}(\bar{0})
+\frac{1}{2}P^{(0)}_{\Delta}(\bar{1})
-\frac{1}{2}P^{(0)}_{\Delta}(\bar{3})
+P^{(0)}_{\Delta}(\bar{4}) \nonumber \\
&=& \Delta_0\left(1+4\Delta_0^2\right)
\sqrt{\frac{P(2,0)P(0)}{P^{(0)}_{\Delta_0}(2,0)P^{(0)}_{\Delta_0}(0)}}
+\Delta_0\left(1-4\Delta_0^2\right)
\frac{P(1)}{P^{(0)}_{\Delta_0}(1)},\label{Exe:trueLRO}
\end{eqnarray}
which indeed is of order $Z$ when $1/2-\Delta_0 \sim Z$, $P(2,0)\sim 1$, 
$P(1)\sim Z$ and $P(0)\sim Z^2$. 
The explicit evaluation of the $Z$ reduction factor is presented in the Appendix, see  
Eq. (\ref{App:ZBCS}). 

Let us now compare the variational energy as given by  
Eq. (\ref{BCS-EG}) with $Z$ of Eq. (\ref{App:BCSdeltasmall}), valid for 
$1/2 - \Delta_0 = \delta \ll 1$, to the energy of a non superconducting 
paramagnetic solution, Eq. (\ref{BCS-EG}) with $\Delta=0$, $Z$ being given 
by Eq. (\ref{App:ZnoBCS}). We find that the Gutzwiller projected BCS 
wave-function has always higher energy by terms roughly of order $Z|T_0|$.

Therefore, even though the Gutzwiller projector is quite 
efficient to transform the huge Hartree-Fock energy cost, namely 
\begin{eqnarray*}
\langle \Phi_{BCS}(\Delta_0) |
\hat{H} | \Phi_{BCS}(\Delta_0) \rangle 
- \langle \Phi_{BCS}(0) |
\hat{H} | \Phi_{BCS}(0) \rangle &=& 
T_0(\Delta_0) - T_0(0) + 2U\Delta_0^2 + 4J_T\Delta_0^2 \\
&& \simeq -T_0(0) + \frac{U}{2} + J_T,
\end{eqnarray*}
into a much smaller one of order $Z|T_0|\sim ZU$ close to the Mott transition, 
yet it is not able to make superconductivity favorable. Namely, the best variational 
metallic solution remains the one described in Section \ref{Mott transition}, with all the 
drawbacks discussed in Section \ref{Weakness}. In conclusion, as we anticipated, the Gutzwiller 
variational approach does not properly describe a mean-field-unlike Mott transition.

In reality we may expect a superconducting phase just before the VB Mott insulator. 
Recently it has been studied a 
model which shares many common features with the present one, namely 
a three-band Hubbard model with inverted Hund's rules \cite{Massimo}, mimicking a 
strongly dynamical Jahn-Teller effect. 
For an average number of electrons per site $n_0=2$, the inverted Hund's rules 
favor, like in our example, a non degenerate singlet on-site configuration. By 
a DMFT calculation, a superconducting instability was discovered just before the singlet 
Mott insulator. However, that instability was found to appear when the quasiparticle residue 
$Z\sim |J|$ (see Ref. \onlinecite{Massimo}). As we discussed at length previously, the simplest 
metallic Gutzwiller wave-function which we have so far considered is unable to reach $Z\sim J$, 
since it becomes disadvantageous with respect to the insulating one 
already at $Z\sim \sqrt{J}$. Therefore we can not exclude that 
superconductivity may occur even in the two-band model we have considered.       

\section{Conclusions}
In this paper we have analysed some peculiar features of the Mott transition 
displayed by a multi-band Gutzwiller variational wave-function (GWF) 
in infinite dimensions. The analysis has been carried out by using the generalized 
GWF introduced in Ref. \onlinecite{Gebhard}, which 
allows a simple analytical treatment in infinite dimensions. Moreover, we have extended 
that wave-function to account for broken-symmetry phases. 

It is usually assumed that the GWF in infinite dimensions 
gives a faithful description of the quasiparticle 
behavior around the Mott metal-insulator transition. We have shown that while 
this belief is partly true for single-band models, it is incorrect for 
multi-band models. In particular, we have identified at least two major failures of the 
GWF across the Mott transition. The first concerns the occupation probability $P(n)$ of 
on-site charge configurations with $n$ electrons different from the 
integer average one $n_0$, which is believed to represent just the quasiparticle occupation 
probability normalized to the quasiparticle residue $Z$.  
The GWF in infinite dimensions predicts $P(n)\sim Z^{|n-n_0|}$ close to the Mott transition, 
while both physical arguments as well as Dynamical Mean Field Theory results 
suggest a $P(n\not = n_0)\sim Z$, 
even in infinite dimensions. This apparently innocuous 
disagreement is instead profound. In fact, the GWF results imply 
that the quasiparticles remain much more strongly interacting      
than what the correct $P(n)\sim Z$ behavior suggests.

Another drawback concerns the Mott transition in the presence of a weak multiplet exchange 
splitting term $J$. Within the GWF, the Mott transition turns into a first order one and 
occurs when the quasiparticle residue $Z\sim \sqrt{|J|/W}$, $W$ being the bare bandwidth, 
much before the quasiparticle gas has had the time to react against $J$. 
This happens because the 
susceptibility to an infinitesimal exchange splitting $J$ diverges at the Mott 
transition for a GWF. In reality, that susceptibility is finite so that the 
Mott transition is either second order or weakly first order, in that case occurring 
when $Z\sim |J|/W$. The main consequence is that the interesting region where the 
metallic hopping-energy gain $\sim Z W$ competes against the exchange $J$ is not even  
accessible by a Gutzwiller wave-function. 
Both the above mentioned shortcomings have the same origin: the inability of the 
GWF to account for virtual processes into unfavorable charge configurations. 

We have then argued, on the basis of a two-band model study, that the GWF is still 
a good variational wave-function in all cases which can be qualitatively described by 
a mean-field theory, but it fails otherwise, as for instance in the case we have 
explicitly analysed where the Mott insulator is a local version of a valence bond insulator. 
There an improvement of the GWF is necessary. 

\begin{acknowledgments}
We acknowledge helpful discussions with Joerg B\"uenemann, Sandro Sorella 
and Federico Becca. We are particularly 
grateful to Claudio Castellani for his un-valuable comments and suggestions. 
This work was partly supported by MIUR COFIN-2001.   
\end{acknowledgments}
  
\appendix
\section{Antiferromagnetic Gutzwiller wave-function for a two-band model}
The best Hartree-Fock wave-function $|\Phi_0(m)\rangle$ 
on a bipartite lattice when the multiplet exchange term favors the spin-triplet 
configuration is the ground state of an Hamiltonian   
\[
\hat{H}_{HF} = \hat{H}_0 + \mu \sum_{i}\sum_{a=1,2} (-1)^i\left( 
n_{i,a\uparrow} - n_{i,a\downarrow} \right),
\]
which describes an antiferromagnetic insulator with order parameter
\[
\langle \Phi_0 |  n_{i,1\uparrow} - n_{i,1\downarrow} | \Phi_0\rangle 
= \langle \Phi_0 |  n_{i,2\uparrow} - n_{i,2\downarrow} | \Phi_0\rangle 
= 2(-1)^i m.
\]
We search for the optimal Gutzwiller wave-function 
\[
|\Psi_G \rangle = \prod_i \hat{P}_{i,G}\, |\Phi_0(m)\rangle,
\]
where for a given sublattice and making use of particle-hole symmetry, 
\begin{eqnarray*}
\hat{P}_{i,G} &=& 
\lambda_0\, 
\left[\, |i,0 \rangle\langle i,0|+|i,4\rangle\langle i,4|\, \right]\\
&&+ \lambda_{1+}\, \left[ \, |i,1;S_z=1/2 \rangle\langle i,1;S_z=1/2| 
+ |i,3;S_z=1/2\rangle\langle i,3;S_z=1/2|\, \right] \\
&&+ \lambda_{1-}\, \left[ \, |i,1;S_z=-1/2 \rangle\langle i,1;S_z=-1/2| 
+ |i,3;S_z=-1/2\rangle\langle i,3;S_z=-1/2|\, \right] \\
&& + \lambda_{2\pm} \, |i,2,\pm \rangle \langle i, 2,\pm |
+ \lambda_{20} \, |i,2,0\rangle \langle i, 2,0 |\\
&& + \lambda_{2t+}\,|i,2;S=1,S_z=1\rangle\langle i,2;S=1,S_z=1| \\
&& + \lambda_{2t-}\,|i,2;S=1,S_z=-1\rangle\langle i,2;S=1,S_z=-1|\\
&& + \lambda_{2t0}\,|i,2;S=1,S_z=0\rangle\langle i,2;S=1,S_z=0|
\end{eqnarray*}
while for the other sublattice $+$ with $-$ interchange. In this particular case 
the Gutzwiller projector remains hermitean since the non-unitary transformation 
$\hat{U}_i$ is diagonal in the above multiplet basis.   

We define the correlated probability distributions for a given sublattice 
\begin{eqnarray*}
P(1,+) &=& P(1,S_z=1/2) = P(3,S_z=1/2),\\
P(1,-) &=& P(1,S_z=-1/2) = P(3,S_z=-1/2),\\
P(2,t+) &=& P(2,S=1,S_z=1), \\
P(2,t-) &=& P(2,S_z=-1),\\
P(2,t0) &=& P(2,S_z=0),
\end{eqnarray*}
and analogously for the uncorrelated ones $P^{(0)}(\dots)$. 
For the other sublattice $S_z \leftrightarrow -S_z$. 

The two conditions (\ref{cond-1}) and (\ref{cond-2}) imply that 
\begin{eqnarray*} 
\lambda_0^2 &=& \frac{P(0)}{P^{(0)}(0)}=\frac{P(0)}{\left(\frac{1}{4}-m^2\right)^2},\\
\lambda_{1+}^2 &=& \frac{P(1,+)}{P^{(0)}(1,+)}
=\frac{P(1,+)}{2\left(\frac{1}{2}+m\right)^3\left(\frac{1}{2}-m\right)},\\
\lambda_{1-}^2 &=& \frac{P(1,-)}{P^{(0)}(1,-)}
=\frac{P(1,-)}{2\left(\frac{1}{2}-m\right)^3\left(\frac{1}{2}+m\right)},\\
\lambda_{2,\pm}^2 &=& \frac{P(2,\pm)}{P^{(0)}(2,\pm)}
=\frac{P(2,\pm)}{2\left(\frac{1}{4}-m^2\right)^2},\\
\lambda_{2,0}^2 &=& \frac{P(2,0)}{P^{(0)}(2,0)}
=\frac{P(2,0)}{\left(\frac{1}{4}-m^2\right)^2},\\
\lambda_{2,t0}^2 &=& \frac{P(2,t0)}{P^{(0)}(2,t0)}
=\frac{P(2,t0)}{\left(\frac{1}{4}-m^2\right)^2},\\
\lambda_{2,t+}^2 &=& \frac{P(2,t+)}{P^{(0)}(2,t+)}
=\frac{P(2,t+)}{\left(\frac{1}{2}+m\right)^4},\\
\lambda_{2,t-}^2 &=& \frac{P(2,t-)}{P^{(0)}(2,t-)}
=\frac{P(2,t-)}{\left(\frac{1}{2}-m\right)^4}.\\
\end{eqnarray*}
The correlated occupation probabilities satisfy the normalization condition
\begin{equation}
2P(0)+2P(1,+)+2P(1,-)+P(2,0)+P(2,\pm)+P(2,t0)+P(2,t+)+P(2,t-)=1,
\label{AF:norm}
\end{equation}
as well as the conservation of the order parameter
\begin{equation}
P(1,+)+P(2,t+)-P(1,-)-P(2,t-)=2m.
\label{AF:cons-m}
\end{equation}
By all the above defined quantities, the variational energy is found to be
\begin{eqnarray}
E_G(m) &=& Z(m)\, T_0(m) + U\left[ 4P(0) + P(1,+) + P(1,-)\right] 
+ 2J_T \, P(2,0)  \nonumber\\
&& + 2J_S\left[P(2,t+)+P(2,t0)+P(2,t-)\right] 
- \left(3J_S+J_T\right)\, P(2,\pm),
\label{AF:EG}
\end{eqnarray}
where
\[
T_0(m) = \langle \Phi_0(m)|\hat{H}_0 | \Phi_0(m)\rangle,
\]
and 
\begin{eqnarray}
\sqrt{Z(m)}&=& \frac{2}{\sqrt{1-4m^2}}\left[ 
\sqrt{P(0)}\left(\sqrt{\frac{P(1,+)}{2}}+\sqrt{\frac{P(1,-)}{2}}\right)\right. \nonumber\\
&& + 
\frac{1}{2}
\left( \sqrt{P(1,+)} + \sqrt{P(1,-)} \right) \left( \sqrt{P(2,\pm)}
+ \sqrt{\frac{P(2,0)}{2}} + \sqrt{\frac{P(2,t0)}{2}}   \right) \nonumber\\
&& \left. + \sqrt{\frac{P(2,t+)P(1,+)}{2}}
+ \sqrt{\frac{P(2,t-)P(1,-)}{2}} \right] \label{AF:Z}
\end{eqnarray}
For any finite $U$ the optimal solution has always $m\not = 0$ due to the nesting 
property. For very large $U$ we expect $m\to 1/2$. In this limit we can neglect all 
$P$'s but $P(1,+)$ and $P(2,t+)$ hence, from Eqs. (\ref{AF:norm}) and (\ref{AF:cons-m}),
\[
P(2,t+) = 4m -1,\;\; P(1,+)= 1-2m,
\]
which implies that 
\[
Z(m) \simeq 2 \frac{4m-1}{1+2m}.
\]
In the same limit the uncorrelated hopping energy has the expression 
\[
T_0(m) \simeq -2\sqrt{M_2\left(2-4m\right)},
\]
where 
\[
M_2 = \int d\epsilon \, \rho(\epsilon)\, \epsilon^2,
\]
is the second moment of the uncorrelated density of states per spin and orbital, 
$\rho(\epsilon)$. Therefore the variational energy as function of the order 
parameter $m$ for $U\gg |T_0|$ is  
\[
E_G(m) \simeq -4 \frac{4m-1}{1+2m} \sqrt{M_2\left(2-4m\right)} 
+ U\,\left(1-2m\right) + 2J_S\, \left(4m-1\right),
\]
and it is optimized by 
\begin{equation}
m \simeq \frac{1}{2} - \frac{M_2}{\left(U-4J_S\right)^2},
\label{AF:m}
\end{equation}
leading to 
\begin{equation}
E_G \simeq 2J_S - \frac{2M_2}{U-4J_S}.
\label{AF:Evar}
\end{equation}

\section{Evaluation of the $Z$-factor for the two-band model} 
The explicit expression of the $Z$ reduction factor in the two-band model of Section 
\ref{A two-band model study} allowing for a superconducting order parameter is, 
through Eq. (\ref{Exe:BCSZ}),
\begin{eqnarray}
\sqrt{Z} &=& 
\lambda_{\bar{0}} \lambda_{\bar{1}} 
\left(\frac{1}{2}+\Delta\right)^3
+\frac{3}{2}\lambda_{\bar{1}} \lambda_{\bar{2}t} 
\left(\frac{1}{2}+\Delta\right)^2
\left(\frac{1}{2}-\Delta\right) \nonumber \\
&+&\lambda_{\bar{1}} \lambda_{\bar{2}\pm} 
\left(\frac{1}{2}+\Delta\right)^2
\left(\frac{1}{2}-\Delta\right)
+\frac{1}{2}\lambda_{\bar{1}} \lambda_{\bar{2}0} 
\left(\frac{1}{2}+\Delta\right)^2
\left(\frac{1}{2}-\Delta\right) \nonumber\\
&+&\frac{3}{2}\lambda_{\bar{3}} \lambda_{\bar{2}t} 
\left(\frac{1}{2}+\Delta\right)
\left(\frac{1}{2}-\Delta\right)^2
+\lambda_{\bar{3}} \lambda_{\bar{2}\pm} 
\left(\frac{1}{2}+\Delta\right)
\left(\frac{1}{2}-\Delta\right)^2\nonumber\\
&+&\frac{1}{2}\lambda_{\bar{3}} \lambda_{\bar{2}0} 
\left(\frac{1}{2}+\Delta\right)
\left(\frac{1}{2}-\Delta\right)^2
+ \lambda_{\bar{3}} \lambda_{\bar{4}} 
\left(\frac{1}{2}-\Delta\right)^3. \label{App:Z}
\end{eqnarray}

If we parametrize the $\lambda$'s according to Eq. (\ref{lambda-parametrization}) 
and make use of (\ref{lambda-BCS}) we find
\begin{eqnarray}
\sqrt{Z} &=& 
\frac{2}{\sqrt{1-4\Delta^2}}\left\{ 
\frac{1}{4}\sqrt{P(0)P(1)}\left[
\sqrt{\frac{P^{(0)}_{\Delta_0}(\bar{0})P^{(0)}_{\Delta_0}(\bar{1})}
{P^{(0)}_{\Delta_0}(0)P^{(0)}_{\Delta_0}(1)}} 
+
\sqrt{\frac{P^{(0)}_{\Delta_0}(\bar{4})P^{(0)}_{\Delta_0}(\bar{3})}
{P^{(0)}_{\Delta_0}(0)P^{(0)}_{\Delta_0}(1)}} \right.\right. \nonumber \\
&+&\left. 
\sqrt{\frac{P^{(0)}_{\Delta_0}(\bar{2},0)P^{(0)}_{\Delta_0}(\bar{1})}
{P^{(0)}_{\Delta_0}(0)P^{(0)}_{\Delta_0}(1)}}
+
\sqrt{\frac{P^{(0)}_{\Delta_0}(\bar{2},0)P^{(0)}_{\Delta_0}(\bar{3})}
{P^{(0)}_{\Delta_0}(0)P^{(0)}_{\Delta_0}(1)}}
\right] \nonumber \\
&+&\frac{1}{4}\sqrt{P(2,0)P(1)}\left[
\sqrt{\frac{P^{(0)}_{\Delta_0}(\bar{0})P^{(0)}_{\Delta_0}(\bar{1})}
{P^{(0)}_{\Delta_0}(2,0)P^{(0)}_{\Delta_0}(1)}}
+
\sqrt{\frac{P^{(0)}_{\Delta_0}(\bar{4})P^{(0)}_{\Delta_0}(\bar{3})}
{P^{(0)}_{\Delta_0}(2,0)P^{(0)}_{\Delta_0}(1)}}
\right] \nonumber \\
&+& \left.
\left[\frac{\sqrt{3}}{4}\sqrt{P(2,t)P(1)} + \frac{\sqrt{2}}{4}\sqrt{P(2,\pm)P(1)}\right]
\left[
\sqrt{\frac{P^{(0)}_{\Delta_0}(\bar{1})}
{P^{(0)}_{\Delta_0}(1)}}
+
\sqrt{\frac{P^{(0)}_{\Delta_0}(\bar{3})}
{P^{(0)}_{\Delta_0}(1)}}
\right]\right\}. 
\label{App:ZBCS}
\end{eqnarray}
When $\Delta=\Delta_0=0$, the above expression reduces to the 
$Z$-factor for a paramagnetic non superconducting solution, namely 
\begin{eqnarray} 
\sqrt{Z} &=& 2\sqrt{P(0)P(1)} + \sqrt{3}\sqrt{P(1)P(2,t)}\nonumber \\
&+& \sqrt{2}\sqrt{P(1)P(2,\pm)} + \sqrt{P(1)P(2,0)}.
\label{App:ZnoBCS}
\end{eqnarray}

The uncorrelated probabilities distributions in the natural basis with 
order parameter $\Delta_0$ (analogous expressions hold for $\Delta$) are  
\begin{equation}
\begin{array}{lcl}
P^{(0)}_{\Delta_0}(\bar{0}) &=& \left(\frac{1}{2}+\Delta_0\right)^4 
\simeq 1 - 4\delta + 6\delta^2, \\
P^{(0)}_{\Delta_0}(\bar{1}) &=& 4\left(\frac{1}{2}+\Delta_0\right)^3
\left(\frac{1}{2}-\Delta_0\right)
\simeq 4\delta - 12\delta^2, \\
P^{(0)}_{\Delta_0}(\bar{2};0) &=& \left(\frac{1}{2}+\Delta_0\right)^2
\left(\frac{1}{2}-\Delta_0\right)^2
\simeq \delta^2,\\
P^{(0)}_{\Delta_0}(\bar{2};\pm) &=& 2\left(\frac{1}{2}+\Delta_0\right)^2
\left(\frac{1}{2}-\Delta_0\right)^2 
\simeq 2\delta^2,\\
P^{(0)}_{\Delta_0}(\bar{2};t) &=& 3\left(\frac{1}{2}+\Delta_0\right)^2
\left(\frac{1}{2}-\Delta_0\right)^2
\simeq 3\delta^2, \\
P^{(0)}_{\Delta_0}(\bar{3}) &=& 4\left(\frac{1}{2}+\Delta_0\right)
\left(\frac{1}{2}-\Delta_0\right)^3 
\simeq 0,\\
P^{(0)}_{\Delta_0}(\bar{4}) &=& \left(\frac{1}{2}-\Delta_0\right)^4 
\simeq 0,
\\
\end{array}
\label{Exe:BCSP_0}
\end{equation}
where the last expressions on the left hand side correspond to 
the limit $\Delta_0 = 1/2 - \delta$ with $\delta\ll 1$. 
The uncorrelated occupation probabilities 
in the original electronic basis are readily obtained by the latter upon inverting 
Eq. (\ref{Pbar-P}):
\begin{equation}
\begin{array}{lcl}
P^{(0)}_{\Delta_0}(0) &=& P^{(0)}_{\Delta_0}(4) 
= \left(\frac{1}{4}+\Delta_0^2\right)^2 \simeq \frac{1}{4}-\delta+2\delta^2,\\
P^{(0)}_{\Delta_0}(1) &=& P^{(0)}_{\Delta_0}(3) 
= 4\left(\frac{1}{16}-\Delta_0^4\right) 
\simeq 2\delta - 6\delta^2,\\
P^{(0)}_{\Delta_0}(2,0) &=& \frac{1}{16} 
+ \frac{3}{2}\Delta_0^2 + \Delta_0^4 
\simeq \frac{1}{2} - 2\delta + 3\delta^2 ,\\
P^{(0)}_{\Delta_0}(2,\pm) &=& 2\left(\frac{1}{4}-\Delta_0^2\right)^2 
\simeq 2\delta^2,\\
P^{(0)}_{\Delta_0}(2,t) &=& 3\left(\frac{1}{4}-\Delta_0^2\right)^2
\simeq 3\delta^2. \\
\end{array}
\label{App:P_0-original}
\end{equation}
In the limit of small $\delta$, by inserting (\ref{Exe:BCSP_0}) 
and (\ref{App:P_0-original}) into (\ref{App:ZBCS}) one finds 
at leading order (recalling that $\Delta \sim Z\ll 1$) 
\begin{eqnarray} 
\sqrt{Z} &\simeq& \sqrt{2}\sqrt{P(0)P(1)}\left(1+\delta\right) 
+ \sqrt{P(1)P(2,0)}
\nonumber \\
&+& \left[\sqrt{\frac{3}{2}}\sqrt{P(1)P(2,t)} \sqrt{P(1)P(2,\pm)}\right]
\left(1+\delta\right) .
\label{App:BCSdeltasmall}
\end{eqnarray}


\begin{thebibliography}{99}
\bibitem{nota} This property follows from the simple fact that the leading terms 
in the variational energy involving the probability $P(n_0\pm m)$, $m>0$, are of the order 
\[
T_0 \sqrt{P(n_0)P(n_0+1)}\sqrt{P(n_0\pm m)P(n_0 \pm m \mp 1)},
\]
for the hopping energy, and 
\[
U P(n_0\pm m),
\]
for the Hubbard repulsion. Comparing the two opposite in sign contributions, one gets 
\[
P(n_0\pm m) \sim \frac{|T_0|}{U} P(n_0 \pm m \mp 1)P(n_0)P(n_0+1),
\]
thus leading to
\[
P(n_0\pm m) \sim Z^m,
\]
since one expects $P(n_0)\sim 1$ and $P(n_0\pm 1)\sim Z$.
\bibitem{nota-b} In reality this is already commonly done 
in numerical simulations when the $t-J$ model is analysed as representative of  
a strong coupling Hubbard model, see Refs. \onlinecite{GrosSC,Sorella,SorellaPRL,Randeria}.
\end{thebibliography}
\end{document}